\documentclass[useAMS]{mn2e}
\usepackage{graphicx}
\newcommand{\bJ}{\bmath{J}}
\newcommand{\bj}{\bmath{j}}

\newcommand{\bB}{\bmath{B}}
\newcommand{\bE}{\bmath{E}}
\newcommand{\bH}{\bmath{H}}
\newcommand{\bD}{\bmath{D}}

\newcommand{\text}[1]{\quad\mbox{#1}\quad}
\newcommand{\spr}[2]{\bmath{#1} \!\cdot\! \bmath{#2}}
\newcommand{\vpr}[2]{\bmath{#1} \!\times\! \bmath{#2}}
\newcommand{\vdiv}[1]{\spr{\nabla}{#1}}

\newcommand{\vcurl}[1]{\vpr{\nabla}{#1}}

\newcommand{\Pd}[1]{\partial_{#1}}

\begin{document}

\title{``Meissner effect'' and Blandford-Znajek mechanism in conductive black hole magnetospheres}


\author[S.S.Komissarov and J.C. McKinney]{S.S.Komissarov$^{1}$ and Jonathan C. McKinney$^{2}$\\
  $^{1}$Department of Applied Mathematics, the University of Leeds,
  Leeds, LS2 9GT, UK. E-mail:serguei@maths.leeds.ac.uk \\
  $^{2}$Institute for Theory and Computation, CfA, 60 Garden St., Cambridge, MA, 02138.  E-mail:jmckinney@cfa.harvard.edu}


\date{Received/Accepted}
\maketitle
                                                                                                
\begin{abstract}
The expulsion of axisymmetric magnetic field from the event horizons
of rapidly rotating black holes has been seen as an astrophysically
important effect that may significantly reduce or even nullify the 
efficiency of the
Blandford-Znajek mechanism of powering the relativistic jets in Active
Galactic Nuclei and Gamma Ray Bursts. However, this Meissner-like effect 
is seen in vacuum solutions of black hole electrodynamics whereas
the Blandford-Znajek mechanism is concerned with plasma-filled
magnetospheres. In this paper we argue that conductivity dramatically
changes the properties of axisymmetric electromagnetic solutions --
even for a maximally rotating Kerr black hole the magnetic field is
pulled inside the event horizon. Moreover, the conditions resulting in
outgoing Poynting flux in the Blandford-Znajek mechanism exist not on
the event horizon but everywhere within the black hole
ergosphere. Thus, the Meissner effect is unlikely to be of interest in
astrophysics of black holes, at least not in the way this has been
suggested so far.  These conclusions are supported by the results of
time-dependent numerical simulations with three different computer codes. 
The test problems involve black holes with the rotation parameter ranging 
from $a=0.999$ to $a=1$.  The pure electrodynamic simulations deal with 
the structure of conductive magnetospheres of black holes placed in 
a uniform-at-infinity magnetic field (Wald's problem) and the 
magnetohydrodynamic simulations are used to study the magnetospheres 
arising in the problem of disk accretion.   
\end{abstract}

                                                                                                
\section{Introduction}

The electromagnetic mechanism of extraction of rotational energy of
black holes by Blandford and Znajek~\shortcite{BZ} is considered as
one of the most promising models for powering the relativistic jets of active
galactic nuclei (AGN), gamma-ray bursts (GRBs), and galactic black
hole binaries. When this mechanism is described in terms of classical
physics the black hole horizon is often compared with a rotating
magnetized conductor \cite{Dam,RuT}. This viewpoint was canonized in
the ``Membrane paradigm'' by Thorne et al.~\shortcite{TPM}. Such a
description puts black holes on the same footing as say the Sun since
the Blandford-Znajek mechanism appears not much different from the
mechanism of magnetic braking by Weber \& Davis~\shortcite{WD}.  In
this respect the so-called ``Meissner effect of black hole
electrodynamics,'' that is the expulsion of magnetic flux from the
horizon of rapidly rotating black holes, seems to undermine the role
of the Blandford-Znajek mechanism in astrophysics as ``the conductor''
becomes unmagnetized.

\begin{figure*}
\includegraphics[width=80mm,angle=-90]{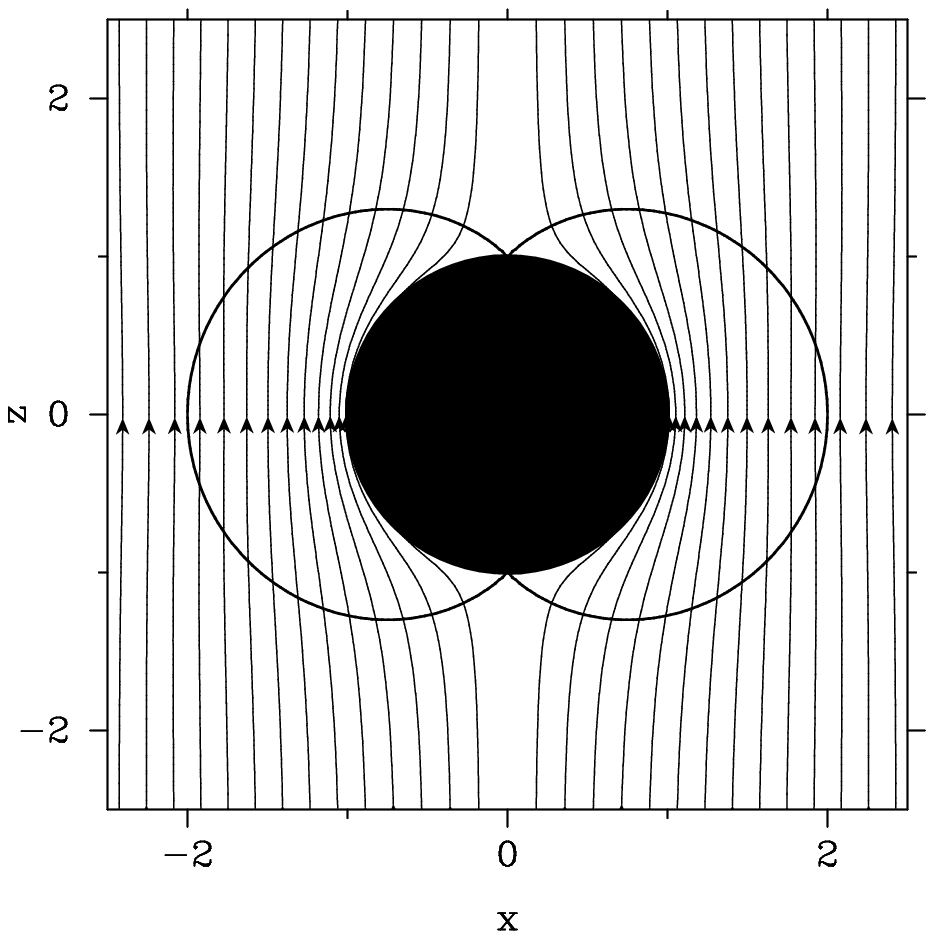}
\includegraphics[width=80mm,angle=-90]{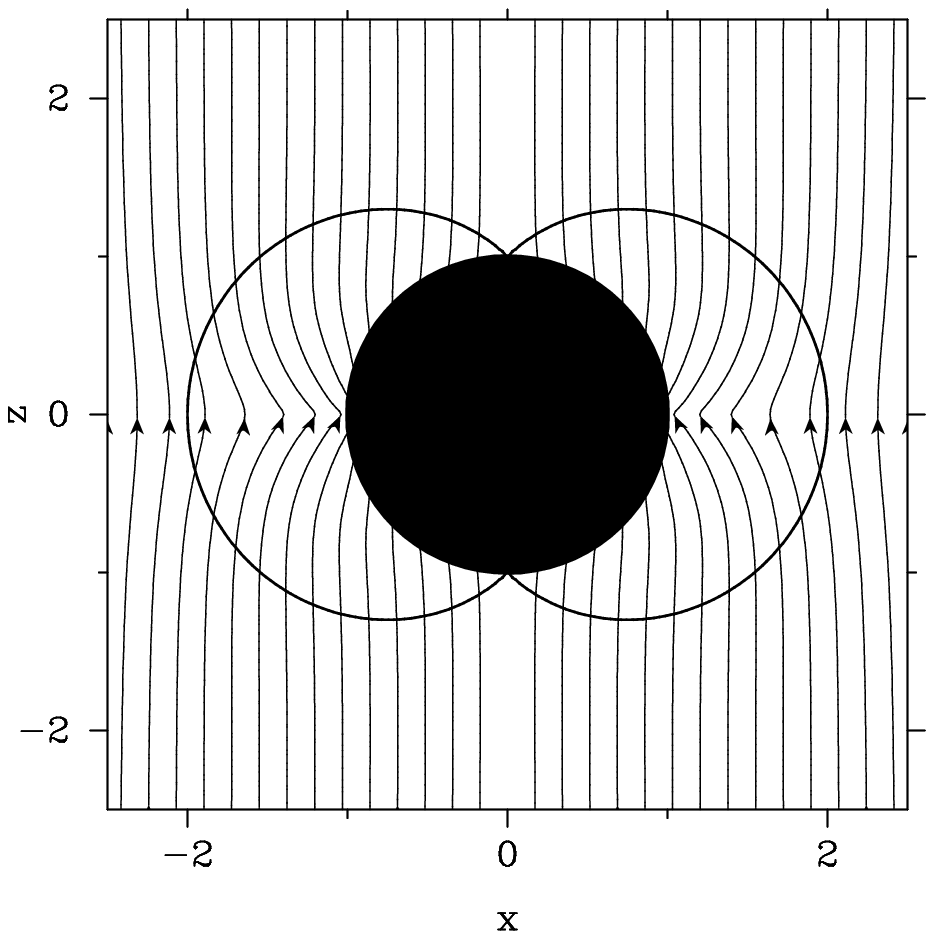}
\caption{ Left panel: Steady-state numerical solution for the vacuum
case.  The magnetic flux is expelled from the event horizon thus
illustrating the ``Meissner effect'' of black hole
electrodynamics. Right panel: Numerical solution for the 
conductive case at time $t=6GM/c^3$. The magnetic flux is pulled back 
onto the horizon. The unit length
corresponds to $GM/c^2$. The thick line shows the black hole
ergosphere ($a=1.0$).} 
\label{c1-2d}
\end{figure*}

Here is the brief history of the Meissner effect.
Wald~\shortcite{W74} found the exact steady-state vacuum solution for
a rotating black hole placed into a uniform-at-infinity magnetic
field aligned with hole's rotational axis. For this solution, King et
al.\shortcite{KLK} computed the total magnetic flux, $\Phi$, threading
the event horizon as a function of hole's rotation parameter, $0\le a
< 1$.  They found that $\Phi \to 0$ as $a \to 1$ and, thus, for
maximally rotating black holes, the magnetic flux is totally expelled
from the black hole horizon (see fig.~\ref{c1-2d}). By its appearance
the phenomenon is similar to the Meissner effect, that is the
expulsion of magnetic field by superconductors.  Later Bi\v c\'ak \&
Jani\v s~\shortcite{BJ} showed that this result held for all
axisymmetric steady-state vacuum solutions and concluded that their
finding has a detrimental effect on the prospects of the
Blandford-Znajek mechanism. Last year this argument was reiterated by
Bi\v c\'ak et al.\shortcite{BKL}.

Recently, there has been impressive progress in numerical methods
for general relativistic magnetohydrodynamics
\cite{KSK,K01,DvH,GMT,K04b,Due,Ant,Ann,SS,AHLN} and a number of groups
have carried out simulations of magnetized flows around Kerr black
holes. None has reported observations of the Meissner-like effect and
so it is natural to ask why. McKinney \& Gammie~\shortcite{MG04}
suggested that this could be due to a relatively small value of
parameter $a$ in their simulations -- indeed, the effect is not strong
unless $a$ is very close to unity.  This explanation has also been put
forward in Bi\v c\'ak et al.\shortcite{BKL}. However there are other
reasons that may be even more important. Indeed, the difference
between the systems of magnetohydrodynamic (MHD) and vacuum
electrodynamics is very significant both in structure and number of
equations. The two physical factors that are accounted for in MHD but
not in vacuum electrodynamics are the conductivity and inertia of
plasma. Both of them have been argued to be of crucial importance for
powering relativistic outflows from black holes \cite{BZ,PC}.

In this paper we show that conductivity alone is sufficient to nullify
the Meissner effect even for maximally rotating black holes. In
particular, we present the results of electrodynamic simulations for
the Kerr black hole with the rotation parameter $a=1.0$ placed in an
aligned uniform magnetic field.  We also present the results of a
general relativistic magnetohydrodynamic (GRMHD) simulation of a
realistic magnetized accretion flow with black hole spin $a=0.999$.
The GRMHD model confirmes that the inclusion of inertia and thermal 
effects does not help to revive the Meissner effect.
  
\section{Electrodynamics of black holes}

In this study we only consider the astrophysically most relevant case
of test fields, that is we assume that the mass-energy of the electromagnetic 
field is too small to effect the curvature of space-time. In our analysis 
we employ the 3+1 formulation of black hole electrodynamics
developed in Komissarov~\shortcite{K04a}.  While the better known
formulation of Macdonald and Thorne~\shortcite{MT82} is adapted to the
Boyer-Lindquist foliation of spacetime that introduces a coordinate
singularity on the event horizon, the formulation in
Komissarov~\shortcite{K04a} is more general and can be used for the
Kerr-Schild foliation as well. This is important because there is no
coordinate singularity in the Kerr-Schild foliation, which therefore
has advantages for both numerical and analytical studies of physical
processes in the vicinity of the event horizon.  In particular, this
allows us to place the inner boundary of the computational domain {\it
inside} the event horizon.  The differential equations of this
formulation are identical to the equation of classical electrodynamics
in matter
\begin{equation}
   \vdiv{B}=0,
\label{divB}
\end{equation}
                                                                                
\begin{equation}
   \Pd{t}\bB + \vcurl{E} = 0,
\label{Faraday}
\end{equation}

\begin{equation}
   \vdiv{D}=\rho,
\label{divD}
\end{equation}
                                                                                
\begin{equation}
   -\Pd{t}\bD + \vcurl{H} = \bJ .
\label{Ampere}
\end{equation}
The purely spacial vectors $\bB,\bD,\bE,\bH$ are related to the Maxwell tensor 
via

\begin{equation}
     B^i= \frac{1}{2}e^{ijk} F_{jk},
\end{equation}
                                                                                                    
\begin{equation}
   E_i = F_{it},
\end{equation}

\begin{equation}
     D^i=\alpha F^{ti},
\label{D1}
\end{equation}
                                                                                                    
\begin{equation}
     H_i =\frac{\alpha}{2} e_{ijk} F^{jk},
\label{H1}
\end{equation}
where $e_{ijk}$ is the Levi-Civita tensor of space. 

The curved space of black hole behaves as an electromagnetically
active medium whose electromagnetic properties are described by the
following constitutive equations

\begin{equation}
     \bE = \alpha \bD + \vpr{\bbeta}{B},
\label{E3}
\end{equation}
                                                                                
\begin{equation}
     \bH = \alpha \bB - \vpr{\bbeta}{D},
\label{H3}
\end{equation}
where $\alpha$ is the lapse function and $\bbeta$ is the shift vector
of the spacetime foliation. Following the general principle of
relativity it is most convenient to describe local physical processes
in the frames of local inertial observers who are instantaneously at
rest in the ``absolute space'' of the foliation (``fiducial
observers'' or FIDOs). Such an observer sees $\bB$ as the local
magnetic field, $\bD$ as the local electric field, and $\rho$ as the
local electric charge density. The local electric current density,
$\bj$, is related to $\bJ$ of the Ampere equation via
    
\begin{equation}
     \bJ=\alpha\bj-\rho\bbeta.
\label{e5}
\end{equation}

Neglecting the inertia of charged particle one can write the
generalized Ohm law as
\begin{equation}
  \bj = \sigma_\parallel \bD_\parallel +
        \sigma_\perp \bD_\perp + \bj_d,
\label{Ohm}
\end{equation}
where 
\begin{equation}
    \bj_d =  \rho \frac{\vpr{\bD}{\bB}}{\check{B}^2}
\label{jdrift2}
\end{equation}
is the drift current, $\sigma_\parallel \bD_\parallel$ is the
conductivity current parallel to the magnetic field and $\sigma_\perp
\bD_\perp$ is the conductivity current perpendicular to the magnetic
field \cite{K04a}.  In a collisionless plasma with a strong magnetic
field the cross-field conductivity is highly suppressed so one can
safely use $\sigma_\perp=0$.  On the contrary, the parallel
conductivity is very large, thus resulting in almost vanishing
parallel component of the electric field, $D_\parallel \ll B$.

\begin{figure*}
\includegraphics[width=80mm,angle=0]{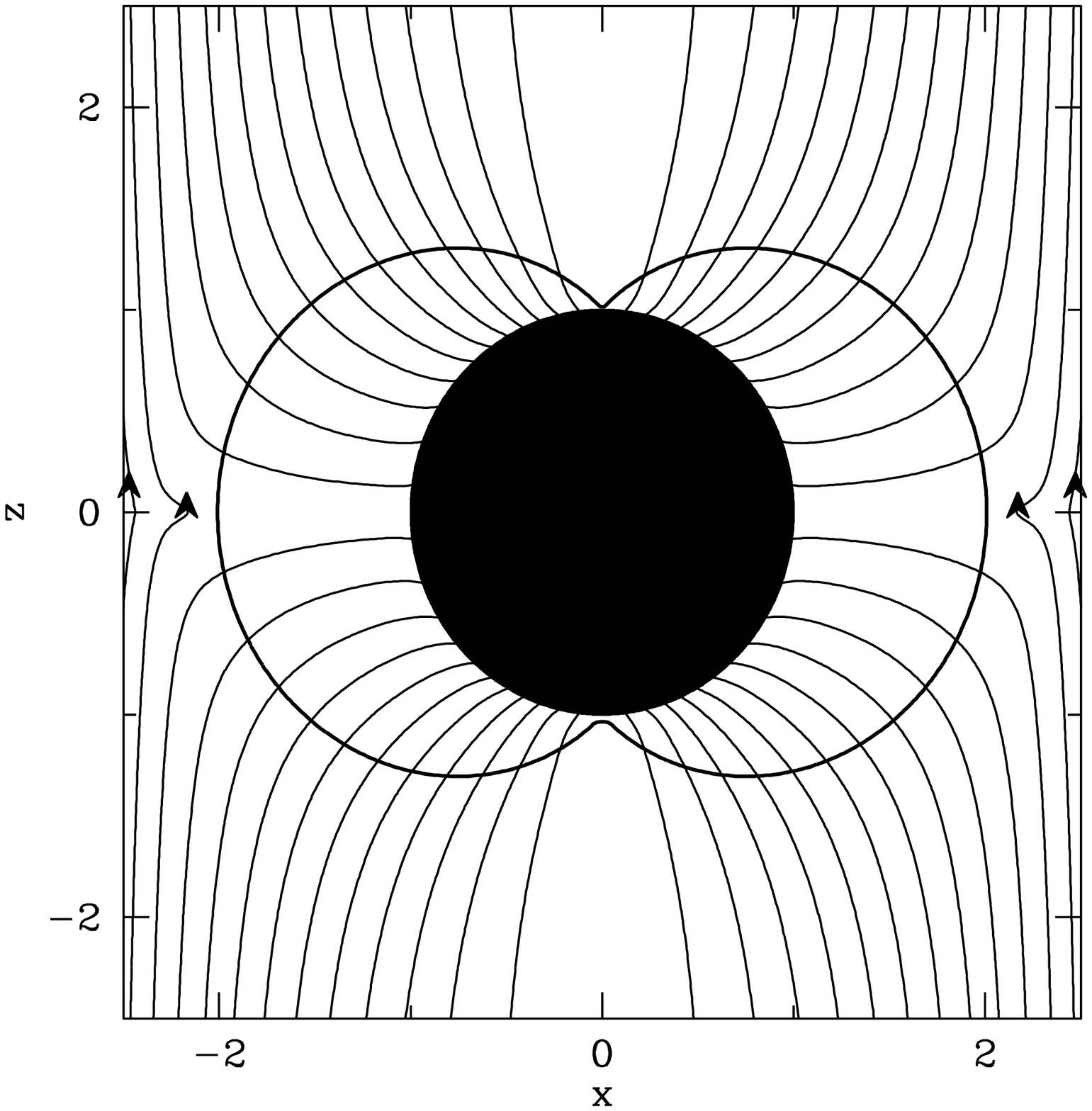}
\includegraphics[width=80mm,angle=0]{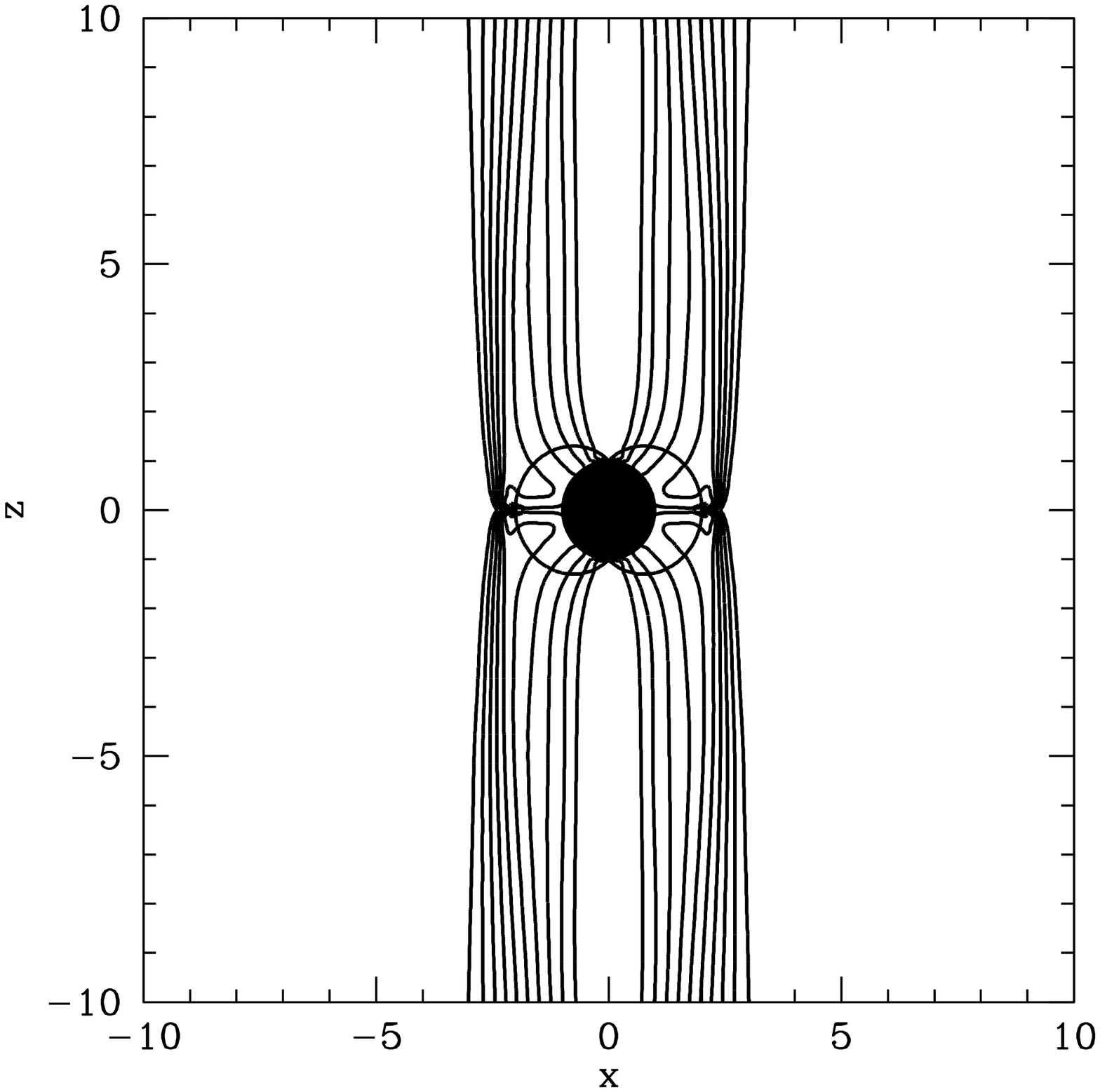}
\caption{Left panel: Similar to the right panel of figure~\ref{c1-2d}
  showing the steady-state numerical solution for the {\it highly}
  conducting case.  All of the flux passing through the ergosphere
  threads the black hole.  Right panel: Same model but shows field
  angular frequency of rotation, where for $x>0$ there are $11$
  contours from $\Omega_F=0$ to $\Omega_F=\Omega_H/2$. This
  demonstrates presence of cylindrically collimated Poynting jet.}
\label{c2-2d}
\end{figure*}

\section{Computer simulations}

In our search for insights into the role of conductivity in the
Meissner effect of black hole electrodynamics, we first consider the 
famous Wald problem \cite{W74}. The steady state solution of vacuum
electrodynamics to this problem describes not only the magnetic field
but the electric field as well. The magnetic part is
\begin{equation}
\begin{array}{rcl}
  B^\phi&=&B_0(g_{\phi r,\theta}+2ag_{tr,\theta})/\sqrt{\gamma}, \\
\label{Wald_B}
  B^r&=&-B_0(g_{\phi\phi,\theta}+2ag_{t\phi,\theta})/\sqrt{\gamma}, \\
  B^\theta&=&B_0(g_{\phi\phi,r}+2ag_{t\phi,r})/\sqrt{\gamma},
\end{array}
\end{equation}
where $t,\phi,r,\theta$ are the Kerr-Schild coordinates, $g_{\nu\mu}$ is
the metric tensor of spacetime, and $\gamma$ is the determinant of the
metric tensor of space. The electric part is

\begin{equation}
\begin{array}{rcl}
  E_\phi&=&0, \\
\label{Wald_E}
  E_r&=&-B_0(g_{t\phi,r}+2ag_{tt,r}), \\
  E_\theta&=&-B_0(g_{t\phi,\theta}+2ag_{tt,\theta}).
\end{array}
\end{equation}
(The components of $\bB$ and $\bE$ are given in the non-normalized
coordinate bases of Kerr-Schild coordinates.)  The emergence of the
electric component in this solution is a particular example of the
electromagnetic activity of curved space-times. This effect is of
great astrophysical importance as it leads to the Blandford-Znajek
process.

The equations of conductive electrodynamics of black holes are more
complicated than those of vacuum electrodynamics and the analytic
approach has not been very successful so far. Fortunately, there are
now a number of numerical techniques that one may use. 
In our study we first used the upwind scheme described 
in Komissarov~\shortcite{K04a} and set $\sigma_\parallel=1/\Delta t$
where $\Delta t$ is the computational time-step. Since $\Delta t$ is
much smaller than the light crossing time of the magnetosphere, the
solution is expected to be fairly close to one with infinite
conductivity, and this is confirmed by the results. 
We used solution (\ref{Wald_B},\ref{Wald_E}) as the initial solution of
our numerical simulations and to make sure that the Meissner effect is
apparent we set $a=1.0$, that corresponds to a maximally
rotating black hole. We carried out two runs: one with vanishing
conductivity, $\sigma_\parallel=0$, and the other with
$\sigma_\parallel=1/\Delta t$, which corresponds to the highly
conductive case (except in current sheets). The left panel of
figure~\ref{c1-2d} shows the steady-state numerical solution of vacuum
equations. One can see that magnetic flux is expelled from the black
hole horizon. In fact, on this plot the numerical solution is
indistinguishable from the exact analytic solution. The right panel of
figure~\ref{c1-2d} shows the conductive case at time
$t=6GM/c^3$. Now the magnetic flux is no longer expelled from the
event horizon but on the contrary the magnetic field lines are
actually attracted to it. By this time the solution starts to develope 
the dissipative equatorial current sheet \cite{K04a}.
Unfortunately, the code cannot handle this kind of current sheets 
where the electric field tends to become stronger than the magnetic field. 
It turns out that the numerical structure of such current sheets in this 
code is controlled by numerical resistivity \cite{K06}.  
Thus this scheme gives reliable results only until the ergospheric current 
sheet is formed.

McKinney~\shortcite{mckinney2006a} constructed a different numerical 
scheme that provides greater control over the influx of energy into 
such currents sheets thus allowing them to evolve to an almosts 
dissipationless state. Figure~\ref{c2-2d} shows the almost steady-state 
solution to the Wald problem with $a=1.0$ obtained with this code 
(a low level unsteady reconnection controlled by numerical resistivity 
occurs in the equatorial current sheet.) 
The left panel of figure~\ref{c2-2d} shows the magnetic field lines 
and the right panel shows their angular velocity , $\Omega_F\equiv
-E_\theta/\sqrt{\gamma}B^r$, where $\gamma$ is the determinant of the
metric tensor of space. This
result is similar to that obtained by Komissarov~\shortcite{K05} for
the GRMHD Wald problem with $a=0.9$ where the reconnection rate in 
the equatorial current sheet is inhibited by the high thermal pressure 
of locally heated plasma. The Blandford-Znajek power from the black hole is
maximally efficient since much of the magnetosphere rotates with the angular
velocity  $\Omega_F =\Omega_H/2$, where $\Omega_H$ is the angular velocity
of the blach hole.  The cylindrically collimated Poynting jet is
driven by both the black hole and the ergospheric disc.

\begin{figure}
\includegraphics[width=80mm,angle=0]{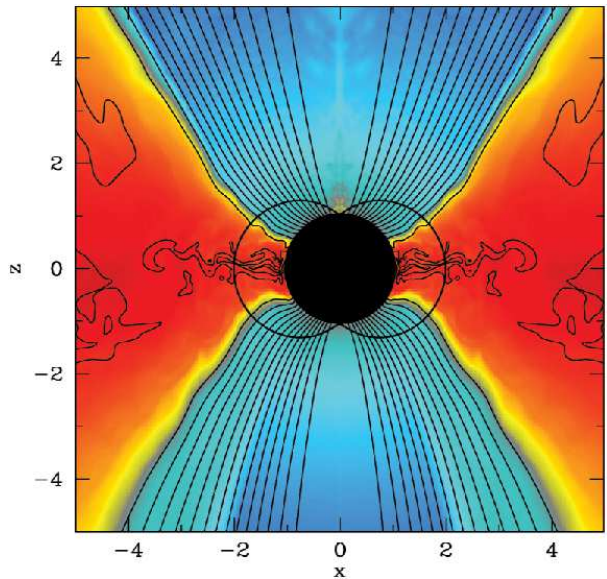}
\caption{Quasi-steady GRMHD magnetized accretion disk solution for 
   a black hole with $a=0.999$ at time $t=2000GM/c^3$ . 
  Contours show the magnetic field lines and the colour 
  image shows logarithm of the rest-mass density.  
  Amount of flux
  threading black hole is qualitatively similar to lower black hole
  spins for similar models.  The Meissner effect is not observed.}
\label{c3-2d}
\end{figure}

Finally, we also use the 
GRMHD code HARM \cite{GMT,mckinney2006b} to confirm the results of 
electrodynamic simulations, to see if the inclusion of plasma inertia and 
thermal pressure makes any difference, and to study the problem in the 
astrophysically more realistic context of disc accretion. Here we examine a GRMHD
solution of a magnetized accretion onto a black hole with 
$a=0.999$ and otherwise similar parameters as described in McKinney \&
Gammie~\shortcite{MG04}.  Figure~\ref{c3-2d} shows the
GRMHD solution for the field geometry and rest-mass density.  We find
that, as in the previous simulations with a slower rotating black 
hole \cite{McK05}, there is no sign of magnetic flux expulsion from 
the event horizon. In the funnel the magnetic geometry is close to the 
monopole assumed in the model by Blandford and Znajek~\shortcite{BZ}. 
Thus, even for very rapidly rotating black holes the Blandford-Znajek 
effect can drive magnetized jets similar to those 
seen in simulations with slower rotating black holes 
\cite{mckinney2006b,HaK06}.

\section{Discussion and Conclusions}

The numerical simulations described here show that the ``Meissner effect'' 
is nullified when high conductivity, typical for the magnetospheres of
astrophysical black holes, is taken into account.  What are the reasons
for this result and how general is it?
In order to answer these questions it is perhaps helpful to speculate
on why one would expect the results of vacuum electrodynamics to be
applicable under the conditions where conductivity is undoubtedly very
high. Consider for example the case of a uniform magnetic field in
flat spacetime. This is an exact steady-state solution of vacuum
Maxwell's equations. Now lets us uniformly fill the space with static
electrically neutral plasma. Obviously, this does not disturb the
equilibrium and the magnetic configuration remains unchanged. The
same applies to any other equilibrium magnetostatic configuration (we
ignore gravity.).

The difference between this case and the case of rotating black holes
is in the fact that strong electric field is necessarily induced in
the vicinity of a black hole when it is placed in vacuum magnetic
field even if the black hole itself has zero electric charge.  In the
particular case of the Wald problem this ``gravito-rotationally induced''
electric field is described by equations
(\ref{E3},\ref{Wald_E}). However, this result is quite generic
\cite{K04a}. Once plasma is introduced in such an electromagnetic
field the charged particles of different signs will move in the
opposite directions bringing about screening of the electric
component. This screening deeply upsets the equilibrium of vacuum
solution as the electric and magnetic fields are tightly coupled via
equations (\ref{E3},\ref{H3}). The effect is strongest in the vicinity
of the event horizon where the magnitudes of electric and magnetic
fields are comparable. Moreover, it has been shown in
Komissarov~\shortcite{K04a} that no static distribution of electric
charge can give full screening of the electric component within the
black hole ergosphere where under this condition the purely poloidal
magnetic field is bound to have lower magnitude than the normal
component of the electric field.  To achieve marginal screening a
poloidal electric current must flow through the ergospheric region
thus strengthening the magnetic field by creating the azimuthal
component that was absent in the vacuum solution.
The hoop stress associated with this component tends to pull the magnetic 
flux back on to the event horizon.

The disscussion of the Meissner effect with connection to the Blandford-Znajek
mechanism in Bi\v c\'ak et al.\shortcite{BKL} highlights the widely spread 
misinterpretation 
of the electromagnetic mechanism by Blandford-Znajek, namely that the energy is 
generated by the rotating event horizon. In fact, it is not the rotation of the 
horizon, or the ``stretched horizon'' \cite{TPM}, that makes possible the 
electromagnetic extraction of the rotational energy of black holes. 
It is much more instructive to consider the space itself as rotating and  
this rotation is particularly strong within the black hole ergosphere where 
it results in unavoidable rotation of plasma in the same sense as the
space (or the black hole). Punsly and Coroniti \shortcite{PCb} were the first 
who correctly and clearly indentified the ergosphere as the key region for the 
magnetic energy extraction and constructed a plausable model of ``inertially 
driven wind'' based on this plasma rotation within the ergosphere. However, in 
their critisim of the 
Blandford-Znajek mechanism they failed to see that this mechanism is also based on 
the extreme conditions existing within the ergosphere and not on the properties
of the event horizon.       
The signature of space rotation, expressed by the shift vector $\bbeta$, is present
outside of the event horizon, where it provides the additional
coupling of the electric and magnetic fields that holds even in 
steady-state configurations (see eqs.\ref{E3},\ref{H3}). 
Screening of the gravito-rotationally induced electric field, that is the field 
generated via this additional coupling, involves generation of stationary
poloidal electric current \cite{K04a} and this current remains nonvanishing even
when a steady-state is reached. It is this current, driven by
the marginally screened ergospheric electric field, that provides for
the energy and angular momentum extraction in the Blandford-Znajek
mechanism and as far as the magnetic flux is not expelled from the
ergosphere, and it is not expelled even in the vacuum solutions, the
electromagnetic mechanism will continue to operate. Thus, even if the                   
Meissner effect did hold in the conductive regime this would not be very 
important for the Blandford-Znajek mechanism.

\section*{Acknowledgments}
SSK was supported by PPARC under the rolling grant
``Theoretical Astrophysics in Leeds.''  JCM was supported by a Harvard
CfA Institute for Theory and Computation fellowship.


\end{document}